\documentclass[11pt]{article}
\input epsf.tex
\newdimen\SaveWidth \SaveWidth=\textwidth
\newdimen\SaveHeight \SaveHeight=\textheight
\advance\SaveWidth by -\textwidth
\advance\SaveHeight by -\textheight

\divide\SaveWidth by 2
\divide\SaveHeight by 2
\advance\hoffset by \SaveWidth
\advance\voffset by \SaveHeight

\def\abs#1{\left| #1\right|}

\let\badcite=\cite
\def\cite{~\badcite}

\def\slashchar#1{\setbox0=\hbox{$#1$}           % set a box for #1
   \dimen0=\wd0                                 % and get its size
   \setbox1=\hbox{/} \dimen1=\wd1               % get size of /
   \ifdim\dimen0>\dimen1                        % #1 is bigger
      \rlap{\hbox to \dimen0{\hfil/\hfil}}      % so center / in box
      #1                                        % and print #1
   \else                                        % / is bigger
      \rlap{\hbox to \dimen1{\hfil$#1$\hfil}}   % so center #1
      /                                         % and print /
   \fi} 
    \def\slashword#1{\setbox0=\hbox{$#1$}        %     set a box for #1
  \dimen0=\wd0                                   %and get its size
   \setbox1=\hbox{/} \dimen1=\wd1                % get size of /
   \ifdim\dimen0>\dimen1                         % #1 is bigger
      \rlap{\hbox to \dimen0{\hfil\bf---\hfil}} %        
      #1                                         %
   \else                                         % / is bigger
      \rlap{\hbox to \dimen1{\hfil$#1$\hfil}}    % so center #1
      /                                          % and print /
    \fi}                                         %

\catcode`@=11
\newdimen\vbigd@men                             % for \vbig

\def\vbig#1#2{{\vbigd@men=#2\divide\vbigd@men by 2%
   \hbox{$\left#1\vbox to \vbigd@men{}\right.\n@space$}}}
\catcode`@=12

\catcode`@=11
\def\citenum#1{\csname b@#1\endcsname}
\catcode`@=12

\def\dofig#1#2{\centerline{\epsfxsize=#1\epsfbox{#2}}}
\begin{document}
\begin{titlepage}
\rightline{LBNL-47091}

\bigskip\bigskip

\begin{center}{\Large\bf\boldmath
Flavor Alignment in SUSY GUTs\footnotemark \\}
\end{center}
\footnotetext{
 This work was supported in part by the Director, Office of Science, Office
of High Energy and Nuclear Physics, Division of High Energy Physics of the
U.S. Department of Energy under
Contracts DE-AC03-76SF00098.
}

\bigskip
\centerline{\bf I. Hinchliffe and N. Kersting}
\centerline{{\it Lawrence Berkeley National Laboratory, Berkeley, CA}}
\bigskip

\begin{abstract}

        A Supersymmetric Grand unified model is  
constructed  based on SO(10)xSO(10) symmetry  in
        which new types of Yukawa matrices couple 
standard and exotic fermions. Evolution of these couplings from the
Grand Unified scale to the electroweak scale causes some of them to be  
        driven to their fixed points.  This solves the  
        supersymmetric  alignment problem and ensures that there are
        no observable flavor changing neutral currents mediated by
        supersymmetric particles.  Fermion hierarchy and neutrino
        mixing  constraints are automatically satisfied in this 
         formalism.

\bigskip        

\end{abstract}

\newpage
\pagestyle{empty}

\end{titlepage}

%%%%%%%%%%%%%%%%%%%%%%%%%%%%%%%%%%%%%%%%%%%%%%%%%%%%%%%%%%%%%%%%%%%%%%
\section{Introduction}
\label{sec:intro}

 Supersymmetry (SUSY),
an as yet undetected but highly promising additional symmetry of
nature, offers a possible solution to the hierarchy problem\cite{hier}.
SUSY still has a tremendous amount of theoretical freedom in
constructing models: the over one 
hundred free parameters in the theory must be fixed.
Models of supersymmetry breaking can be invoked to reduce the number
of arbitrary parameters. In gauge mediated models \cite{GMSB1,GMSB2} the
pattern of SUSY breaking parameters leads to a low energy theory that
has no dangerous flavor changing neutral currents(FCNCs). Models based on
gravity mediation of SUSY breaking\cite{supergrav},
 reduce parameter space considerably, but have no 
 compelling reasons for the absence of 
 flavor changing neutral currents.

It is possible that some of the low energy parameters 
are determined independent of their values at the Grand Unified Theory (GUT) scale\cite{dyn1,dyn2}. 
This happens if the 
renormalization group equations (RGEs), governing the evolution of 
the parameters of the theory as
one lowers the energy scale, drive some of the parameters to
 their fixed points. The fundamental values at the GUT scale are then
irrelevant.
 A dynamic reduction of parameter 
space in this manner  
automatically solves the FCNC constraint\cite{alignment}, also 
known as the ``SUSY Flavor Alignment'' problem.

In this paper, we propose a new model which is ``complete'' in the 
sense of being both
SUSY and a GUT and illustrates how the idea
might work in practice.
We revive the Pati-Salam Left-Right idea\cite{pati_salam} in
an $SO(10)_L \times SO(10)_R$ framework, enlarging the particle 
spectrum of the MSSM to include 
an extra three generations of exotic particles, too heavy 
to be seen at the electroweak scale.
 As pointed out previously\cite{alignment}, it is not
possible for the standard Yukawas coupling ordinary matter to be 
close to their fixed points at the weak
scale: the fixed point solutions are all $O(1)$, yet $m_u << m_t$,
implying a large hierarchy of Yukawa couplings. 
 The main function of the exotics
is to provide Yukawa couplings to ordinary (super)matter 
which do run to their fixed points, 
leaving the usual Yukawas free.
In addition, these exotics provide a natural setting for a seesaw effect\cite{seesaw}
 in the mass matrices, establishing 
the observed mass hierarchies in the quark and lepton sectors. 

The layout of this paper is as follows: in Section
\ref{sec:susyflavor} 
we review the 
SUSY Flavor Alignment problem and its resolution via the RGE
fixed-point running; 
 Section \ref{sec:model} introduces the specifics of the model we
 propose, 
including the symmetry-breaking
structure, the particle representations, and the couplings that 
emerge from the superpotential;
 we discuss the vacuum structure of the theory in 
Section \ref{sec:model2} and demonstrate how the
gauge coupling constants evolve from the GUT scale to the weak scale; 
in Section \ref{sec:constpre}
we verify fermion mass hierarchies and mixings and make some general 
comments regarding signatures
of this model in future experiments.  Section \ref{sec:concl} 
summarizes the main conclusions of 
this paper.

\section{SUSY Flavor Alignment}
\label{sec:susyflavor}
        The MSSM for 3 generations of matter, represented by the superpotential
\begin{equation}
\label{superpot}
{\bf W} =  {\bf y_u} \overline{U} Q H_2 +  {\bf y_d} \overline{D} Q H_1
  + {\bf y_e} \overline{E} L H_1 + \mu H_1 H_2
\end{equation}
with the chiral matter superfields for quarks and leptons $\overline{U}$,
 $\overline{D}$, $Q$, $\overline{E}$, $L$, and Higgs $H_{1,2}$
 charged under $SU(3)_C \times SU(2)_L \times U(1)_Y$ 
 assigns the 3x3 Yukawa matrices ${\bf y_{e,d,u}}$ to the coupling between particles of different
 generations. If 
supersymmetry were an unbroken symmetry of nature, then, after the
 Higgs fields obtain VEVs, O(3) rotations in generation space on the quark
 superfields diagonalize the mass terms and gives rise to the usual CKM matrix in the 
(s)quark-(s)quark gauge interactions; experiments would report the same CKM matrix for quarks 
and squarks. However, since SUSY is broken at an energy $M_{SUSY}$, the
theory admits generic 
soft-breaking Yukawa terms in the Lagrangian below this scale:
\begin{equation}
\label{softsusy}
 \begin{array}{rrr}

{\cal L} \supset &  ( {\bf a_u} \overline{U}
         Q H_2 +  {\bf a_d} \overline{D} Q H_1
  + {\bf a_e} \overline{E} L H_1 )_{scalar} + (M_3 {\tilde g} {\tilde g } +
                                M_2 {\tilde W} {\tilde W} + 
        M_1 {\tilde B} {\tilde B} )  \\ 
    &  + (\sum_{X_i=Q,\overline{u},\overline{d},\overline{e}} 
    X^{\dagger}_i {\bf m^2_i} X_i + \sum_{i=1,2} m^2_{H_i} H_i^* H_i
 + b \mu H_1 H_2)_{scalar} + h.c. & \\
\end{array}
\end{equation}

where (${\tilde g}$, ${\tilde W}$, ${\tilde B}$) are the superpartners of the Standard Model gauge
bosons and where {\it scalar} implies that only the scalar component of each 
superfield is used.
Note that the SUSY-breaking parameters  ${\bf a_{u,d,e}}$ and ${\bf m^2_i}$ are matrices with 
{\it a priori} unknown structure.   
The squarks have therefore additional 
sources of inter-generational mixing beyond that for the quarks. At the EW scale the  
quarks' and squarks' mass terms are not in general diagonalizable by the same rotations and 
we have separate ``CKM'' matrices. Experiments such as neutral $K$-meson mixing constrain the squark-CKM matrix to be a 
fraction of a percent deviant from 
the quark CKM  \cite{ckms}. Since such a high degree of alignment between independently chosen 
matrices is unlikely, the MSSM is by itself insufficient to naturally account for
low-energy flavor structure; this is termed the `SUSY Alignment Problem'.

The simplest solution to this problem is to impose a universality on the soft-breaking 
terms, forcing the SUSY-breaking Yukawas ${\bf a_{e,d,u}}$ to be proportional to the
corresponding SUSY Yukawas ${\bf y_{e,d,u}}$. This is useful for computational purposes yet
it lacks any physical motivation. Some models of SUSY breaking such as gauge-mediation 
\cite{GMSB1,GMSB2} have automatic flavor alignment.

There is however an elegant solution \cite{alignment}: the alignment arises inevitably from
the RGEs running to their fixed points. The reason why this occurs is simple: first
 instead of the explicit soft-breaking terms in the Lagrangian as in (\ref{softsusy}) , we can redefine 
SUSY-breaking terms as spurions\cite{spurions} which get 
$\theta$-space dependent vacuum expectation values (VEVs). These VEVs are interpreted as
 parameters in 
the superpotential and gauge couplings after making the substitutions: 
\begin{equation}
\label{thetas}
 \begin{array}{lll}
        g^2  & \rightarrow &  g^2 (1 + M\theta^2 + M\overline{\theta}^2
                + 2 M^2 \theta^2 \overline{\theta}^2)  \\

        {\bf y}^{ijk}  & \rightarrow & {\bf y}^{ijk} - {\bf a}^{ijk} \theta^2
        +{1\over2}({\bf y}^{njk}{\bf m^2}^{i}_{n} + {\bf y}^{ink}{\bf m^2}^{j}_{n}
                + {\bf y}^{ijn}{\bf m^2}^{k}_{n})\theta^2 \overline{\theta}^2     \\
\end{array}
\end{equation}
Here the $\theta^2\overline{\theta}^2$-dependent terms contribute to the effective action starting
at the 1-loop level. 
Now let the Yukawa couplings all run down to their fixed points; since 
the RGEs are gauge-dominated, we expect the fixed point solutions for the Yukawas to have 
a structure that is only a function of the gauge couplings. In particular, each fixed-point 
solution should factorize into a 3x3 matrix of constants and a gauge-dependent piece since 
gauge interactions do not depend on flavor:   
\begin{equation}
\label{fpsol}
{\bf y}_{f.p.} =  {\bf C}\times \psi ({g_i}^2)
\end{equation}
When we switch to the basis where the Yukawa matrix above is diagonal, 
performing rotations $Q \rightarrow \Omega Q$ on the left-handed quark fields and 
separate rotations on each right-handed field,
we diagonalize ${\bf y}_{f.p.}{{\bf y^\dagger}_{f.p.}}$ as well, obtaining from (\ref{fpsol})
\begin{equation}
\label{fpsol2}
{\bf y'}_{f.p.}{{\bf y'^\dagger}_{f.p.}} =  {\bf C'}\times \psi ({g_i}^2)
\end{equation}
where ${\bf y'}_{f.p.}{{\bf y'^\dagger}_{f.p.}} =
 \Omega  {\bf y}_{f.p.}{{\bf y^\dagger}_{f.p.}} {\Omega^\dagger}$   and
${\bf C'} =  \Omega {\bf C }  {\Omega^\dagger}$ , both of which are diagonal in flavor space.
By using the symmetry-breaking rule (\ref{thetas}), we can evaluate  
${\bf y}{\bf y^\dagger}$ after SUSY breaking and quark-field rotation at the fixed point:
\begin{equation}
\label{fpsol3} 
{\bf y}_{f.p.}{{\bf y^\dagger}_{f.p.}} \rightarrow {\bf y'}_{f.p.}{{\bf y'^\dagger}_{f.p.}} 
                -{\bf a'}_{f.p.}{{\bf y'^\dagger}_{f.p.}} \theta^2 + O(\overline{\theta}) 
\end{equation}
As (\ref{fpsol2}) and (\ref{fpsol3}) must agree at each order in $\theta$, we obtain
\begin{equation}
\label{fpsol4}
 {\bf a'}_{f.p.}{{\bf y'^\dagger}_{f.p.}} = - {\bf C'}\times \int{\psi ({g_i}^2)d^2\theta}
\end{equation}
By design ${\bf C'}$ and ${\bf y'}$ are diagonal, and so ${\bf a'}$ is too (likewise
one may check  ${\bf y}{\bf m^2}$): SUSY breaking matrices 
are simultaneously diagonalizable with SUSY Yukawas on the fixed point. This solves the SUSY 
alignment problem if the amount of RGE running is 
sufficient to let the fixed-point properties emerge near the electro-weak scale. 
In this scenario the overwhelming arbitrariness of the soft-SUSY breaking matrices vanishes:
${\bf a}$ and ${\bf m^2}$ are
proportional to ${\bf y}$, the proportionality constant being some function of the various
 coefficients
in the RGEs.

However, since the RGEs' fixed points are functions of $O(1)$ coefficients we would expect the 
Yukawa fixed points to all be of order unity, giving a nearly degenerate mass spectrum.
Since 
huge mass hierarchies exist at the EW scale ($m_t/m_u \approx 10^4$ \cite{4th}), the 
simplest implementation does not work. Taking a cue from \cite{alignment},
we propose a model to address this.

\section{The $SO(10)_L \times SO(10)_R$ Framework }
\label{sec:model}

 A model with only MSSM fields may be immediately ruled out because the Yukawa couplings are
much too small to have reached $O(1)$ fixed points. The simplest variant is 
to let the alignment mechanism discussed above proceed through exotic Yukawas 
running. Schematically the total Yukawa structure in the low energy Lagrangian would 
appear as:
\begin{equation}
\label{yukawas}
{\cal L} ~~ \supset ~~ 
\left( {\bf f} ~~ {\bf f_H} \right)
\left( \begin{array}{c|c}
                        {\bf y} v_L &  {\bf y_1} v_1  \\
                        \hline
                     {\bf y_2} v_2 &  {\bf Y} \Lambda \\
\end{array} \right)
\left( \begin{array}{c}
        {\bf f^c} \\
        {\bf {f_H}^c} \\
\end{array} \right)
\end{equation}
 Here ${\bf f},{\bf f_H}$ denote vectors of standard and exotic fermions, respectively, which need not share
the same dimensionality for the general mechanism to operate. The VEV of the Higgs coupling to ${\bf y}$ is
essentially fixed to be $ v_L \approx m_W$ from experiment, but the magnitudes of the other VEVs $ v_1, v_2, 
\Lambda $ are free parameters. 
 Provided ${\bf y_{1,2}}$ run to their fixed points an alignment can be obtained.
 A suitable hierarchy among the VEVs, $v_L << v_1,v_2 << \Lambda$ may provide
a seesaw solution of the fermion hierarchy problem.
A possible hierarchy of VEVs puts the usual  $O(m_W)$ in the  ${\bf y}$ corner, 
 $O(10~TeV)$ VEVs in the  ${\bf y_x}$ entries, and perhaps  VEVs as large as $O(10^{17} GeV)$
in the  ${\bf Y}$ corner.

Possibly the most elegant realization of  (\ref{yukawas}) involves a left-right symmetry 
which, upon breaking, yields three generations of exotic $SU(2)_L$ singlets which 
couple to themselves through $ {\bf Y}$ and to MSSM fields through ${\bf y_x}$ \cite{minimal}.
The naive breaking pattern 
$SU(5)_L \times SU(5)_R \rightarrow SU(3)_C  \times  SU(2)_L \times SU(2)_R \times U(1)
\rightarrow (standard~~model) $ cannot readily accomodate the low-energy values of the
gauge couplings: $sin^2 \theta_W$ invariably turns out too small\cite{minimalfail}.

        To address this failure we suggest that the low-energy $SU(3)\times U(1)$ color-electromagnetic symmetry is 
the survivor of the following breaking chain:
\begin{equation}
\label{symm}
\begin{array}{c}
                SO(10)_L \times SO(10)_R \\
\downarrow \; \mathtt{M_{GUT}} \\
SU(5)_L \times U(1)'_L \times SU(5)_R \times U(1)'_R \\
\downarrow \; \mathtt{M_{L}} \\
SU(3)_L \times SU(2)_L \times U(1)_L \times SU(5)_R \\
\downarrow \; \mathtt{M_{R}} \\
SU(3)_L \times SU(2)_L \times U(1)_L \times SU(3)_R \times SU(2)_R \times U(1)_R \\
\downarrow \; \mathtt{\Lambda_{LR}} \\
SU(3)_{C} \times U(1)_{L+R}  \times SU(2)_L  \times SU(2)_R \\
\downarrow \; \mathtt{v_{R}} \\
SU(3)_{C} \times U(1)_Y \times SU(2)_L \\
\downarrow \; \mathtt{v_{L}} \\
SU(3)_{C} \times U(1)_{EM}\\
\end{array}
\end{equation}
with the appropriate energy scales noted at the breaking points. In this scheme the scales
in the Yukawa matrix (\ref{yukawas}) correspond as $v_L \sim  \mathtt{v_{L}}$, 
$v_{1,2} \sim  \mathtt{v_{L,R}}$, and $ \Lambda \sim  \mathtt{\Lambda_{LR}}$.
The scale of SUSY breaking cannot be much larger than $1$ TeV as we assume 
SUSY solves the hierarchy problem.

All of the standard model quarks and leptons are unified at the GUT scale  
in three generations of $SO(10)_L \times SO(10)_R$ spinor representations
\begin{equation}
\label{reps}
  \begin{array}{lcccr}
        \{ \; \chi_L({\bf 16};{\bf1}) & \oplus & \chi_R ({\bf1};\overline{{\bf16}}) \; \} & 
        \times \; \; 3 \\
  \end{array}
\end{equation}
The Higgs particles necessary for each stage of the symmetry breaking scheme 
(\ref{symm}) are contained in the bispinor and tensor representations
\begin{equation}
\label{reps2}
  \begin{array}{lcr}
          \overline{\Phi}({\bf16};\overline{{\bf16}}) & \oplus &
                 \Phi(\overline{{\bf16}};{\bf16})  \\
          \Delta_L({\bf10};{\bf1}) & \oplus &  \Delta_R({\bf1};\overline{{\bf10}})  \\
          \Delta'_L({\bf10};{\bf1}) & \oplus &  \Delta'_R({\bf1};\overline{{\bf10}})  \\
          \Delta''_L({\bf10};{\bf1}) & \oplus &  \Delta''_R({\bf1};\overline{{\bf10}})  \\
          \Theta_L({\bf45};{\bf1}) & \oplus &  \Theta_R({\bf1};{\bf45})  \\
          \Theta'_L({\bf45};{\bf1}) & \oplus &  \Theta'_R({\bf1};{\bf45})  \\
        & \oplus & \Theta({\bf45};{\bf45})  \\
  \end{array}
\end{equation}
These choices of representations are not unique, but they are 
the minimal set necessary to avoid fine-tuning in the superpotential, as we demonstrate
later.
 
Whereas the Higgs sector is quite flexible in this model, the choice of three
 matter generations in (\ref{reps}) is more or less fixed, being
tightly constrained by searches for a fourth neutrino at collider energies \cite{4th}.
For ease of discussion in the following, we list in Table~\ref{breaktable} (See Appendix)
 the types of particles present at each stage in the symmetry chain (\ref{symm}).

First we note how the standard model fermions are embedded in the $SO(10)$ rep's: each $16$ 
of  $SO(10)$ decomposes into a $1 \bigoplus \overline{5} \bigoplus 10$ of $SU(5)$. 
Each collection of 16 states, as in conventional $SO(10)$ theories\cite{so10}, 
represents quarks $U$ and $D$ (12 states) and leptons $L$ and  $E$ (3 states) plus a
right-handed neutrino superfield, $N$ (1 state).
Let us show $\chi_L \bigoplus \chi_R$, for example, in more suggestive $SU(5)$ 
language: 
\begin{equation}
\label{su5reps1}
\begin{array}{cc} 
\psi_L = \left( 
                \begin{array}{c} \overline{d}_{H} \\
                                 \overline{d}_{H} \\
                                 \overline{d}_{H} \\
                                   e_L \\
                                   -\nu_L \\ \end{array}     \right)
 & \psi_R = \left( 
                \begin{array}{c} d_{H} \\
                                d_{H} \\
                                d_{H} \\
                                   \overline{e}_R \\
                                   -\overline{\nu}_R \\ \end{array}     \right) \\
\Psi_L = \left(
          \begin{array}{ccccc}
          0   & \overline{u_H} & -\overline{u_H} & -u & -d \\
          -\overline{u_H} & 0 & \overline{u_H} & -u & -d \\
         \overline{u_H} & -\overline{u_H} & 0 & -u & -d \\
        u & u & u & 0 & -\overline{e}_{H} \\
        d & d & d & \overline{e}_{H} & 0 \\ 
        \end{array}  \right)
& \Psi_R = \left(
          \begin{array}{ccccc}
          0   & u_{H} & -u_{H} & -\overline{u_R} &  -\overline{d_R}\\
          -u_{H} & 0 & u_{H} &  -\overline{u_R} &  -\overline{d_R} \\
         u_{H} & -u_{H} & 0 &  -\overline{u_R} &  -\overline{d_R} \\
        \overline{u_R} & -\overline{u_R}  & -\overline{u_R}  & 0 & -e_{H} \\
         -\overline{d_R} & -\overline{d_R}  & -\overline{d_R} & e_{H} & 0 \\ 
        \end{array}  \right) \\
\\
N_L & N_R \\
\end{array}
\end{equation}
All of the fields with an $H$-subscript will acquire masses of order $\Lambda_{LR}$; they
are ``exotic'' particles which are $SU(2)_L$ singlets at the weak scale. The
neutrinos $N_{L,R}$ are likewise heavy, having masses $O(M_{GUT})$. The other particles
in (\ref{su5reps1}) form one standard model generation, including a right-handed 
neutrino, $\nu_R$. Note that the 'mirror symmetry' 
is not the usual one between standard fields and exotics\cite{mirror}.

$SO(10)_L \times SO(10)_R$ breaks to 
$SU(5)_L \times U(1)'_L \times SU(5)_R \times U(1)'_R$ by the singlet components of
the Higgs sector (the ({\bf1};{\bf1}) pieces of $\Phi,\overline{\Phi}$ )
acquiring a VEV. The $SU(5)_{L,R}$ symmetries are broken by the 
 $SU(5)$ adjoint fields contained in $\Theta_{L,R},\Theta'_{L,R}$ and $\Theta$ 
when they get VEVs of the type:
\begin{equation}
\label{sigmavevs}
 <\Theta_x> = i~\sigma_2 \times \left( \begin{array}{ccccc}
                a & 0 & 0 & 0 & 0 \\
                0 & a & 0 & 0 & 0 \\
                0 & 0 & a & 0 & 0 \\
                0 & 0 & 0 & b & 0 \\
                0 & 0 & 0 & 0 & b \\ \end{array}  \right)
\end{equation} 
where the particular values of $a,b$ will depend on $x\in \{L,L',R,R'\}$: for the breaking of 
 $SU(5)_{L,R}$,  $a \sim M_{L,R}$, whereas $b$ may be zero or non-zero 
(see the vacuum structure discussion below in \ref{sec:model2}). 
The Dimopoulos-Wilczek mechanism \cite{dw} for splitting doublets from triplets 
requires no fine-tuning among the VEVs and parameters in the superpotential and ensures that
the colored components ($i=1,2,3$) of all the $SU(5)$ fundamental and 
anti-fundamental Higgs ( ${H_{x}}_{i},\phi_i$, see Table \ref{breaktable} ) get masses $O(M_{L,R})$ whereas the $b$'s are chosen to
make the weak components of $H_L,\overline{H_L},H_R,\overline{H_R},\phi_L,\phi_R$ 
remain light.

When the  $SU(5)_R$ symmetry breaks, it combines with the  $SU(3)_L$  to form a 
vector  $SU(3)_{L+R}$ which remains unbroken all the way down to low energies; this we
interpret as color. At the same time, the $U(1)$'s in the 
theory combine to form a vector $U(1)_{L+R}$.  This vector $U(1)$ is $U(1)_{B-L}$ for
the standard model fields as 
is evident from the charge assignments in Table \ref{breaktable}; the
exotic particles which are absent in the low-energy regime have different charges.
The above breaking to  $SU(3)_C \times U(1)$ is achieved by VEVs of
the fields $\omega, \overline{\omega} , \Omega , \overline{\Omega}$ at $O(\Lambda_{LR})$;
 these themselves acquire masses and decouple from the theory, leaving the uncolored
components of $H_L, \overline{H_L},H_R, \overline{H_R}$ (all from the $\Delta_{L,R}$) and
$\phi_{L,R}$ (from $\Phi$) as light degrees of freedom.   After the breaking 
of $SU(2)_R$, 
the  $SU(2)_L$ symmetry breaks as in the MSSM with  $H_L, \overline{H_L}$
serving as $H_{1,2}$ in (\ref{superpot}).

\subsection{Some Minimal Requirements}
        The relative sizes of the VEVs in (\ref{symm}) are not completely arbitrary.
Simple theoretical and phenomenological considerations allow us to put constraints
on the magnitudes of $v_L,v_R,\Lambda_{LR}, M_L,M_R$, and $M_{GUT}$.
        First of all there must be enough RGE running for the Yukawas to reach
fixed points and for the alignment to work. Since
it is crucial that the Yukawas $\bf {y_{1,2}}$ 
in (\ref{yukawas}) which mix exotic and standard generations run to 
their fixed points, we must ensure that the running, roughly between
 $\Lambda_{LR}$ and $M_{GUT}$, is large enough. Of course in order for the RGEs
 themselves to remain predictive,
we must also ensure that no coupling goes nonperturbative ({\it i.e.} $g \geq 1$)
in this regime; typically this forces the GUT scale itself to respect an upper bound 
just below the Planck scale. 
        There are also important phenomenological constraints which all GUTs must satisfy: masses
of new gauge bosons and proton decay.
 That no experiment has yet detected a $W_R$, or
right-handed version of the $W$-boson, sets a lower bound on its mass of
about 1 TeV  \cite{W_R1, W_R2}. Since we expect $m_{W_R} \approx v_R$, we must have $v_R > 1$ TeV.
 As for proton decay in this model, the usual intermediate gauge boson channels
are closed: 
left- and right-handed quarks are embedded in completely different $SU(5)$
 representations. Proton decay
can proceed through colored Higgs' exchange; specifically, $qqql$ operators arise
from the exchange of $H_x,{\phi}_x$. However these 
operators will be suppressed by the masses of the colored components of these fields 
which are of order $M_L$ and $M_R$. These masses
can actually be as low as $10^{10}$ GeV\cite{protondecay} and still satisfy the 
proton lifetime constraint $\tau_p > 5.5\times 10^{32} yr$ \cite{protonlife}.

Finally, for the exotic fields to have escaped detection, we impose the constraints
on the mass of a fourth generation quark\cite{4th}, giving $\Lambda_{LR} \geq 200$ GeV. This is a lower bound, 
but in fact
we will find it advantageous in Section \ref{sec:model2} to consider much larger  $\Lambda_{LR}$.
        
Altogether, we may list the various phenomenological constraints on the VEVs:
\begin{equation}
\label{constraints}
\begin{array}{ccl}
        M_{GUT} & \leq &  10^{18} GeV \\
        M_{GUT}/v_R & \geq & 10^{10}  \\
        M_{L,R} & \geq & 10^{10} GeV \\
        \Lambda_{LR},v_R & \geq & 10^3 GeV \\
        v_L & \approx & 10^2 GeV \\     
\end{array}
\end{equation}

\subsection{Superpotential and Yukawa Structure}
One way to make the  Dimopoulos-Wilczek mechanism operate effectively \cite{dw}
 is to impose a $Z_3$ symmetry
 on the superpotential at the GUT scale with charges
\begin{equation}
\begin{array}{ll}
0: ~~~& \{\chi_{L,R} ~~  \Delta_{L,R}  ~~  \Theta ~~ \Phi ~~  \overline{\Phi} \} \\
1: ~~~&  \Theta_{L,R} \\
2: ~~~&  \{\Delta'_{L,R} ~~  \Delta''_{L,R} ~~  \Theta'_{L,R}\} \\
\end{array}
\end{equation}
The most general $SO(10)_L \times SO(10)_R$ superpotential is then 
\begin{equation}
W = W_Y + W_H
\end{equation}
where
\begin{equation}
\label{so10sp}
\begin{array}{l}
W_Y =  {\bf \lambda_1} \chi_L \chi_R \Phi  +  {\bf \lambda_2} \chi_L^2 \Delta_L  +  
{\bf \lambda_3} \chi_R^2 \Delta_R \\
\\
\begin{array}{cccccccc}
W_H = &  \lambda_4 \Phi \overline{\Phi}  &  +  & \lambda_5 \Phi \overline{\Phi} \Theta 
 &  + &  \lambda_6 \Delta_L \Theta_L  \Delta'_L  & +  & \lambda_7  \Delta_L \Theta_L  \Delta''_L\\
 &  +  \lambda_8 \Delta'_L \Theta'_L  \Delta''_L &  + & \lambda_9 \Delta_R \Theta_R  \Delta'_R 
      &  + & \lambda_{10}  \Delta_R \Theta_R  \Delta''_R    
      &  + &  \lambda_{11} \Delta'_R \Theta'_R  \Delta''_R \\
 & +  \lambda_{12} \Theta^2 & + & \lambda_{13} \Theta_L  \Theta'_L 
       & + &  \lambda_{14} \Theta_R  \Theta'_R\\
\end{array} 
\\
\end{array}
\end{equation}

In the above the $ {\Delta'}_{L,R},{\Delta''}_{L,R}$ provide a coupling
which splits the doublets from the triplets in the physical $ {\Delta}_{L,R}$ fields which propagate
at low energies ($\Delta_L^2 \Theta_L$ for example vanishes
by the antisymmetry of the ${\bf 45}$). The  $ {\Delta'}_{L,R},{\Delta''}_{L,R}$
 are otherwise inert in the theory as we can arrange
for them to have masses of $O(M_{L,R})$ taking $a,b \sim M_{L,R}$ as discussed earlier (see (\ref{sigmavevs}) ff.).

Under $SU(5)_L\times SU(5)_R \times U(1)^2$, the fields decompose as follows:
\begin{equation}
\label{su5reps}
\begin{array}{lll}
\chi_L & \longrightarrow & \psi_L \oplus \Psi_L \oplus N_L \\
\chi_R & \longrightarrow & \psi_R \oplus \Psi_R \oplus N_R \\
\Delta_L & \longrightarrow & H_L \oplus \overline{H}_L \\
\Delta_R & \longrightarrow & H_R \oplus \overline{H}_R \\
\Delta'_L & \longrightarrow & H_{L'} \oplus \overline{H}_{L'} \\
\Delta'_R & \longrightarrow & H_{R'} \oplus \overline{H}_{R'} \\
\Delta''_L & \longrightarrow & H_{L''} \oplus \overline{H}_{L''} \\
\Delta''_R & \longrightarrow & H_{R''} \oplus \overline{H}_{R''} \\
\Phi & \longrightarrow & \phi_0 \oplus  \phi_L \oplus  \phi_R \oplus \omega \oplus \Omega
                        \oplus \sigma_1 \oplus \sigma_2 \oplus \sigma_3 \oplus \sigma_4 \\
\overline{\Phi} & \longrightarrow & \overline{\phi}_0 \oplus  \overline{\phi}_L \oplus  \overline{\phi}_R
                 \oplus \overline{\omega} \oplus \overline{\Omega}
                        \oplus \overline{\sigma}_1 \oplus \overline{\sigma}_2 \oplus \overline{\sigma}_3
                 \oplus \overline{\sigma}_4 \\
\Theta_L &  \longrightarrow & \Theta_{L1}  \oplus  \Theta_{L2} \oplus \Theta_{L3} \oplus \Sigma_L \\
& \cdot\\
&\cdot\\
&\cdot\\
\end{array}
\end{equation}
in accord with the branching rules ${\bf 16} \longrightarrow {\bf \overline{5}} \oplus {\bf 10} \oplus {\bf 1}$
 , ${\bf 10} \longrightarrow   {\bf 5} \oplus {\bf \overline{5}}$,
 ${\bf 45} \longrightarrow   {\bf 1} \oplus {\bf 10} \oplus  {\bf \overline{10}} \oplus {\bf 24}$,
 and their conjugates. 

In this notation, the effective Yukawa terms just below $M_{GUT}$ in (\ref{so10sp}) become
\begin{equation}
\label{su5yuk}
\begin{array}{lll}
W_Y  & = & {\bf\lambda_1} (\psi_L \psi_R \omega +  \Psi_L \Psi_R \Omega
                        + f(\sigma_i) ) \\
& & +  {\bf\lambda_2} (\psi_L \Psi_L \overline{H}_L + \Psi_L \Psi_L \overline{H}_L) \\
& & +  {\bf\lambda_3} (\psi_R \Psi_R \overline{H}_R + \Psi_R \Psi_R \overline{H}_R) \\  
\end{array}
\end{equation}

Group theory indices are implicit; $\Psi_L \Psi_R \Omega$ means
${\Psi_L} ^{\alpha\beta} {\Psi_R} _{\alpha'\beta'}  \Omega ^{\alpha' \beta'}_{\alpha \beta}$,
for example.
Generation indices on  ${\bf\lambda_{1,2,3}}$ are
likewise hereafter suppressed. 
The first two ${\bf \lambda_1}$-terms give large masses to the exotic particles after 
$\omega$ and  $\Omega$ get VEVs. Note in particular that
the exotic neutrinos $N_{L,R}$ get a GUT-scale mass given by the VEV of the $SU(5)_L \times SU(5)_R$
singlet $\phi_0$. Terms dependent on the $\sigma_i$-fields will not be phenomenologically 
relevant in the remainder of this
study; we choose the vacuum structure of the theory (discussed in Section \ref{sec:model2} below)
 to guarantee this. 

The  $\lambda_{2,3}$-terms of (\ref{su5yuk}) play the role of the couplings ${\bf y_x}$ in 
(\ref{yukawas}) responsible for driving the alignment in the RGEs, with the uncolored components of 
the $H_{L,R}$ fields getting VEVs at $v_{L,R}$ respectively, 
{\it e.g.} $ {\bf \lambda_2} \Psi_L \Psi_L \overline{H}_L$ mixes $\overline{{u_H}}$ and $u$. 
Because the right-handed sector mixes the opposite chirality combination,
{\it e.g.} $ {\bf \lambda_3} \Psi_R \Psi_R \overline{H}_R$ 
mixes $\overline{{u}}$ and $u_H$, we see that
 ${\bf y_1} v_1 \ne {\bf y_2} v_2$ in general and one of $v_{1,2}$ is necessarily $O(m_W)$.
As long as  ${\bf y} << {\bf y_{1,2}}$ the alignment will work; 
small standard Yukawas  ${\bf y}$ are guaranteed in this model both by symmetry 
and the choice of vacuum (see (\ref{omegavevs}) below).

\section{The Model:Vacuum Structure}
\label{sec:model2}

\subsection{Minimizing the Scalar Potential}
The scalar potential receives contributions from the superpotential  
\begin{equation}
\label{scalarpot}
\begin{array}{cccccccc}
W_H = &  \lambda_4 \Phi \overline{\Phi}  &  +  & \lambda_5 \Phi \overline{\Phi} \Theta 
 &  + &  \lambda_6 \Delta_L \Theta_L  \Delta'_L  & +  & \lambda_7  \Delta_L \Theta_L  \Delta''_L\\
 &  +  \lambda_8 \Delta'_L \Theta'_L  \Delta''_L &  + & \lambda_9 \Delta_R \Theta_R  \Delta'_R 
      &  + & \lambda_{10}  \Delta_R \Theta_R  \Delta''_R    
      &  + &  \lambda_{11} \Delta'_R \Theta'_R  \Delta''_R \\
 & +  \lambda_{12} \Theta^2 & + & \lambda_{13} \Theta_L  \Theta'_L 
       & + &  \lambda_{14} \Theta_R  \Theta'_R\\
\end{array} 
\end{equation}
For this choice of
representations the D-term contributions to the scalar potential from $\Delta$'s 
and $\Phi, \overline{\Phi}$ vanish in the limit where
the VEVs from each conjugate pair of fields are equal, whereas those for
 $\Theta$'s vanish according to their symmetry
structure (\ref{sigmavevs}).
Because we have a theory which remains supersymmetric all the way down past $v_R$, the 
minimization of the effective potential implies
\begin{equation}
\label{mineq}
\frac{\partial W}{\partial \eta} = 0  ~~~~~~ 
\eta \supset \{scalars\} 
\end{equation}

 Since in 
(\ref{scalarpot})  the $SO(10)^2$ 
symmetry is already broken by the  $(SU(5) \times U(1))^2$ singlet
 fields $\phi_0, \overline{\phi_0}$,
these fields are effectively non-propagating below $M_{GUT}$. The next symmetries
 to break are $SU(5)_{L,R}$: as 
noted earlier in Section \ref{sec:model}, the $\Theta$, $\Theta_{L,R}$, and $\Theta'_{L,R}$
 fields get diagonal VEVs 
as in (\ref{sigmavevs}), breaking the $SU(5)$s to the subgroup structure
$SU(3)\times SU(2)\times U(1)$ and splitting the colored triplets from the doublets in
the $H$'s and  $\phi$'s. We will obtain four sets of light ($O(v_{L,R})$) doublets 
 $H_L, \overline{H_L},H_R, \overline{H_R},\phi_{L,R},\overline{\phi_{L,R}}$, four sets of heavy doublets
 $H_{L',L''}, \overline{H_{L',L''}},H_{R',R''}, \overline{H_{R',R''}}$, and
all color triplets heavy for the VEV structure
\begin{equation}
\label{hvevs}
\begin{array}{l}
 <\Theta_{L,R}> = i~\sigma_2 \times \left( \begin{array}{ccccc}
                M_{L,R} & 0 & 0 & 0 & 0 \\
                0 & M_{L,R}  & 0 & 0 & 0 \\
                0 & 0 & M_{L,R}  & 0 & 0 \\
                0 & 0 & 0 & v_{L,R} & 0 \\
                0 & 0 & 0 & 0 & v_{L,R} \\ \end{array}  \right) \\
 <\Theta'_{L,R}>, <\Theta> = i~\sigma_2 \times \left( \begin{array}{ccccc}
                M_{L,R} & 0 & 0 & 0 & 0 \\
                0 & M_{L,R}  & 0 & 0 & 0 \\
                0 & 0 & M_{L,R}  & 0 & 0 \\
                0 & 0 & 0 & M_{L,R} & 0 \\
                0 & 0 & 0 & 0 & M_{L,R} \\ \end{array}  \right) \\
\end{array}
\end{equation} 
Below $M_R$ we find the following VEV structure accommodates a minimum:
\begin{equation}
\label{omegavevs}
\begin{array}{ll}
 <\omega> = \left( \begin{array}{ccccc}
                \Lambda_{LR} & 0 & 0 & 0 & 0 \\
                0 & \Lambda_{LR} & 0 & 0 & 0 \\
                0 & 0 & \Lambda_{LR} & 0 & 0 \\
                0 & 0 & 0 & 0 & 0 \\
                0 & 0 & 0 & 0 & 0 \\ \end{array}  \right)
 &  <\overline{\omega}> = \left( \begin{array}{ccccc}
                \Lambda_{LR} & 0 & 0 & 0 & 0 \\
                0 & \Lambda_{LR} & 0 & 0 & 0 \\
                0 & 0 & \Lambda_{LR} & 0 & 0 \\
                0 & 0 & 0 & 0 & 0 \\
                0 & 0 & 0 & 0 & 0 \\ \end{array}  \right) \\
\end{array}
\end{equation}
\begin{equation}
\label{Omegavevs}
\begin{array}{l}
 <\Omega^{\alpha' \beta'}_{\alpha \beta}> = 
  <\overline{\Omega}^{\alpha' \beta'}_{\alpha \beta}> = \Lambda_{LR} ~~~~~~~~~~ for \\
     ~~~~~~~~~~~~~~~~~~~~~~
        \left( \begin{array}{cc} \alpha' &   \beta' \\
                                \alpha &  \beta \\
        \end{array} \right)
        \subset \left( \left( \begin{array}{cc} 1 & 2 \\
                                   1 & 2 \\
        \end{array} \right) ,
         \left( \begin{array}{cc} 1 & 3 \\
                                   1 & 3 \\
        \end{array} \right) ,
         \left( \begin{array}{cc} 2 & 3 \\
                                   2 & 3 \\
        \end{array} \right) ,
         \left( \begin{array}{cc} 4 & 5 \\
                                   4 & 5 \\
        \end{array} \right) \right) \\
\\
   $~~~~~~~~~~~= 0 $ ~~~~otherwise  \\ 
\end{array}
\end{equation}
In the above it should be remembered that  
$<\Omega^{\alpha' \beta'}_{\alpha \beta}> = -<\Omega^{ \beta' \alpha'}_{\alpha \beta}> =
<\Omega^{ \beta' \alpha'}_{ \beta\alpha }>$ (and likewise for $ <\overline{\Omega}>$)
since both $\Omega$ and $\overline{\Omega}$ are antisymmetric $SU(5) \times SU(5)$ tensors.
Also, factors of $O(1)$ may multiply the above VEVs without upsetting the minimization. There is 
a vacuum degeneracy but we assume the particular vacuum in (\ref{Omegavevs}). 
Besides breaking the theory to a vector $SU(3) \times U(1)$, the VEVs above 
give masses to the exotic fields ($u_H,d_H,e_H$) of order $\Lambda_{LR}$. 
This is a crucial 
ingredient of the seesaw mechanism which we discuss in more detail in Section \ref{sec:constpre}.

Since SUSY is assumed to be broken below $v_R$ we must include the SUSY-breaking terms
when breaking $SU(2)_L$. This is
a standard exercise in the MSSM, where the scalar potential takes the form 
\begin{equation}
\label{mssmscalpot}
V = (|\mu|^2 + m_{H_1}^2)|H_{1}^{0}|^2 +  (|\mu|^2 + m_{H_2}^2)|H_{2}^{0}|^2
        - b H_{1}^{0} H_{2}^{0} -  b^* H_{1}^{0*} H_{2}^{0*}
+ {1\over8} (g^2 + g'^2)(|H_{1}^{0}|^2 - |H_{2}^{0}|^2)^2
\end{equation}
We ensure that a similar potential arises in our model at $v_L$:
 simply rename $H_{1,2}$ to $H_L, \overline{H_L}$ in (\ref{mssmscalpot}). The gauge couplings $g,g'$ are understood 
as usual to be the coupling constants
of $SU(2)_L \times U(1)_{Y}$. The low energy spectrum will match the MSSM except now FCNCs are naturally suppressed
and a mass hierarchy is automatic (as discussed in Section \ref{sec:constpre}).

\subsection{Beta Functions and Running Couplings}
Since we are working with a GUT, we require that 
the couplings unify at $M_{GUT}$. 
Each VEV is associated with a threshold in the RGEs where the group symmetry 
and number of propagating particles change, so the running serves as an 
indirect constraint on the values of  $v_L,v_R,\Lambda_{LR}, M_L,M_R$, 
and $M_{GUT}$.
We use 1-loop $\beta$-functions at each
stage of the supersymmetric theory where the symmetry group takes the form 
$G \times G_1 \times \cdots \times G_n$:
\begin{equation}
\label{betaeqn}
\beta = \frac{{g}^3}{16 {\pi}^2} \left( \sum_{i} {C_2}(R_i)d_1(R_i)\cdots d_n(R_i)
                        - 3 C_1(G) \right) + o(g^5)
\end{equation}
Here $C_2(R_i)$ is the index of the irreducible representation $R_i$, defined as
\begin{equation}
C_2(R) \delta ^{ab} \equiv Tr[{\bf T_R}^a {\bf T_R}^b]
\end{equation}
for generators ${\bf T_R}$ in the representation  $R$, $C_1(G)$ is 
 $C_2(R_{adj})$ with $R_{adj}$ the adjoint rep of $G$, and $d_i(R_j)$ is the dimension of
the representation  $R_i$ in the algebra $G_i$. We use the version of (\ref{betaeqn})
with broken SUSY at energies below $1~$TeV: 
\begin{equation}
\label{betaeqn2}
\beta_{n.S.} = \frac{{g}^3}{16 {\pi}^2} \left( {2\over3} \sum_{f} {C_2}(R_f)d_1(R_f)\cdots d_n(R_f)
+ {1\over3} \sum_{s} {C_2}(R_s)d_1(R_s)\cdots d_n(R_s)
                        - {11\over3} C_1(G) \right) +  o(g^5)
\end{equation}
with now separate sums over fermionic and scalar rep's. At each characteristic energy scale, 
the number of particle rep's to account for in (\ref{betaeqn}) or (\ref{betaeqn2}) changes:
below $M_{L,R}$, the $\Sigma$'s, $\sigma$'s, $\Delta'$'s and $\Delta''$'s decouple;
 below $\Lambda$, the $\omega$, $\overline{\omega}$,
$\Omega$, and $\overline{\Omega}$ decouple; finally below $v_R$, 
$\phi_R,\overline{\phi}_R, H_R,\overline{H}_R$ 
decouple. The detailed derivation of the $\beta$-functions above for the present model appear
in the Appendix.
Here we present the result of a sample running in Figure \ref{rgefig} below, with the choices 
of parameters 
\begin{equation}
\label{params}
\begin{array}{l}
\alpha_{GUT}^{-1} = 10 \\
M_{GUT} \ge 10^{16} GeV \\
M_L = 10^{16} GeV  \\
M_R = 10^{15} GeV \\
\Lambda_{LR} = 5 \cdot 10^{12} GeV \\
v_R = 5~TeV \\
\end{array}
\end{equation}

\begin{figure}[t]
\dofig{5.00in}{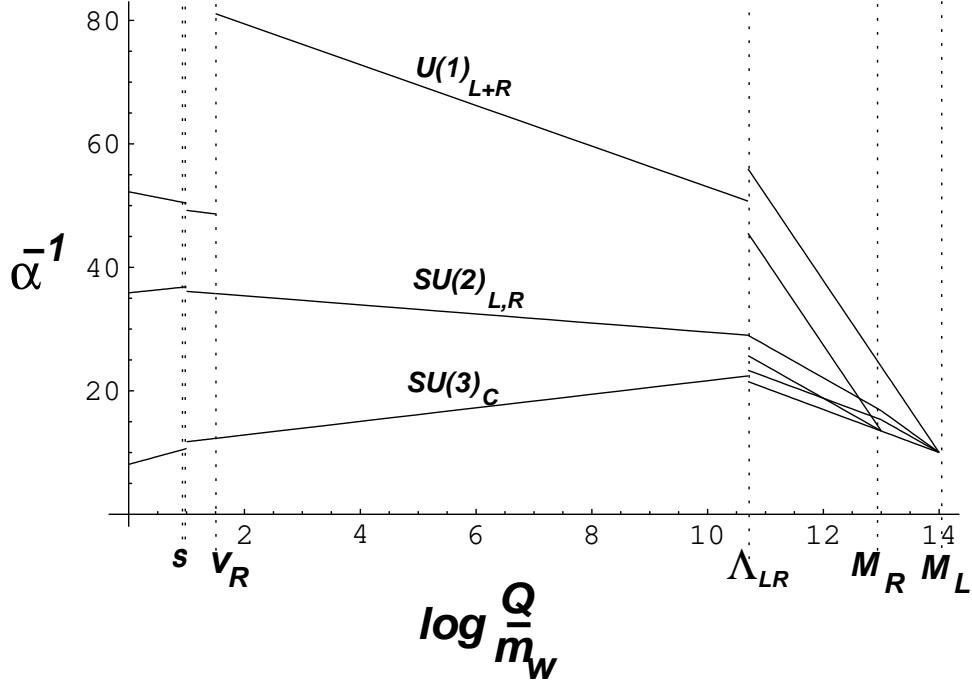}
\caption{One-Loop Running of the Coupling Constants {\it with the choice of parameters
in (\ref{params}). In the interval $(M_R,M_L)$, the runnings are for(top to
bottom) $U(1)_L$, $SU(2)_L$, $SU(3)_L$, and  $SU(5)_R$; 
just above $\Lambda_{LR}$, 
the couplings are for (top to bottom)  $U(1)_L$,  $U(1)_R$,  $SU(2)_L$,
 $SU(2)_R$, $SU(3)_L$, and  $SU(3)_R$. The runnings are computed
with non-SUSY $\beta$-functions below the scale  $s \sim 1~TeV$. Threshold
effects are neglected.} }
\label{rgefig}
\end{figure}

These choices satisfy the
constraints (\ref{constraints}) and are not fine-tuned. From the figure it is clear that
the coupling constants at the weak scale are in rough agreement with the experimental values,
$\alpha_{s}^{-1} = 8.33$, $\alpha_{L}^{-1} = 29.69$, and $\alpha_{Y}^{-1} = 58.8$ \cite{4th};
We expect the agreement to be much better after accounting for particle threshold
effects and using two-loop RGEs, as studies of similar models
confirm \cite{twoloops}. For the sake of our discussion, however, the qualitative agreement
between the weak-scale couplings predicted from this model and the experimental values is
sufficient to consider the parameter choices (\ref{params}) as 'typical'.

\subsection{Low Energy Fermion Mass Matrices}

Putting together the Yukawa couplings in (\ref{yukawas}), the Higgs' VEVs  (\ref{omegavevs}), 
and the constraints  (\ref{params}), we obtain a seesaw-type mass matrix for the fermions at
energies $< v_L$. Interestingly, the matrix structure exhibits a dichotomy between quarks/charged leptons
and neutrinos. For the quarks and charged leptons the mass matrix takes the general form
\begin{equation}
\label{massmatrix}
\left(
\begin{array}{c}
        f_1 \\
        f_2 \\
        f_3 \\
        {f_H}_1 \\
        {f_H}_2 \\
        {f_H}_3 \\
\end{array}
\right)^{\dagger}
\left(
\begin{array}{ccc|ccc}
I & \cdot & \cdot & II & \cdot & \cdot \\
\cdot & o(<<~v_L) & \cdot & \cdot &  o(v_L){\bf\lambda_2} & \cdot \\
\cdot & \cdot & \cdot & \cdot & \cdot & \cdot  \\
\hline
III & \cdot & \cdot & IV & \cdot & \cdot  \\
\cdot & o(v_R) {\bf\lambda_3} & \cdot & \cdot & o(\Lambda_{LR}) {\bf\lambda_1} & \cdot  \\
\cdot & \cdot & \cdot & \cdot & \cdot & \cdot  \\
\end{array}
\right)
\left(
\begin{array}{c}
        {f_1}^c \\
        {f_2}^c \\
        {f_3}^c \\
        {{f_H}_1}^c \\
        {{f_H}_2}^c \\
        {{f_H}_3}^c \\
\end{array}
\right)
\end{equation}
Here the $f$'s represent standard model fermions, whereas the ${f_H}$'s are heavy exotic fields, all of
which come in three generations. Strictly speaking, the entries in Quadrant I of (\ref{massmatrix}) 
are zero in this model, but one can relax this with a  VEV structure slightly differing from 
(\ref{omegavevs}) or with radiative effects. Provided that these entries are small, our results are
essentially unchanged. We have noted the matrix structure in the other quadrants
in accordance with the notation of the superpotential (\ref{su5yuk}). The exotics $u_H,d_H$ and $e_H$
acquire heavy masses in Quadrant IV of $O(\Lambda_{LR})$.

The neutrino sector exhibits a completely different mixing structure. In the basis
$\psi_{\nu} \equiv ({\bf \nu_L},{\bf \nu_R}^c, {\bf N_L}, {\bf N_R}^c)$ the $12 \times 12$-mixing matrix 
has the form
\begin{equation}
\label{neutmatrix}
\overline{\psi}_{\nu}^c ~
\left(
\begin{array}{cc|cc}
 \cdot & \cdot &  o(v_L){\bf\lambda_2} &  o(v_L){\bf\lambda_1} \\
 \cdot & \cdot &  o(v_R){\bf\lambda_1} &  o(v_R){\bf\lambda_3} \\
\hline
 o(v_L){\bf\lambda_2} &  o(v_R){\bf\lambda_1} & 0 &  o(M_{GUT}){\bf\lambda_1} \\
 o(v_L){\bf\lambda_1} &  o(v_R){\bf\lambda_3} & o(M_{GUT}){\bf\lambda_1} & 0 \\
\end{array}
\right)
~ \psi_{\nu}^c
\end{equation}
There are several interesting features of the above structure:
\begin{itemize}
\item  the seesaw for the neutrinos is more severe than for the quarks and charged leptons, since 
        $v_L v_R / M_{GUT} << v_L v_R / \Lambda_{LR}$. This results in much smaller masses for
        neutrinos than for charged leptons. This is in agreement with
        the tiny mass limits on neutrinos \cite{neutrino_mass}.
\item  the flavor mixing in the leptonic sector can be quite different from that in quark sector
\item  the physical neutrinos will be Majorana particles
\end{itemize} 
In the next section we will investigate more fully the range of parameters in the above mass matrices
which can accomodate the known masses and mixings of the standard generations.

\section{Constraints and Predictions}
\label{sec:constpre}

\subsection{Fermion Masses and Neutrino-Mixing}

The low energy manifestation of this model is essentially contained in the structure of the 
mass matrices  (\ref{massmatrix}), (\ref{neutmatrix})
  and the mixing matrices  derived from them in the presence of weak interactions.
In this section we will demonstrate how the parameters of 
the model can accomodate the empirical bounds on these two sets of measurements.

The most recent determination of quark masses \cite{4th} gives
\begin{equation}
\label{qmasses}
\begin{array}{ll}
m_u = 1 ~to~ 5 ~ MeV &  m_d = 3 ~to~ 9 ~ MeV \\
m_c = 1.15 ~to~ 1.35 ~ GeV   &  m_s = 75 ~to~ 170 ~ MeV\\
m_t = 174.3 ~\pm~ 5.1 ~ GeV   &  m_b = 4 ~to~ 4.4 ~ GeV \\ 
\end{array}
\end{equation}
Our model must be able to replicate these hierarchies (as well as the leptonic ones). Fortunately the seesaw 
mechanism is a natural feature of the model, as is evident in  the structure of (\ref{massmatrix}). 
For a typical parameter set as in  (\ref{params}),  the extreme range of the quark masses in (\ref{qmasses})
 is replaced by two much smaller imposed disparities: first, some $O(1)$ inhomogeneity of the small Yukawas
in Quadrant I of  (\ref{massmatrix}), and secondly, some favoring of the heavier quarks to mix more with
the extremely heavy exotics. We reserve numerical details for the Appendix, where our results indicate that such
 mixing may occur at a level sufficient to drive up the heavy quark masses, yet
still not contribute to FCNCs, as we now discuss.  

Constraints on fermion mixing involve both quarks and neutrinos. Both types of mixing 
at present have fairly well measured bounds, yet have completely opposite structures: quarks are
 observed to mix minimally with each other, yet recent experiments seem to suggest  that neutrinos
prefer to mix in a maximal fashion\cite{neutexp}.

Quark mixing is straightforward: upon diagonalizing 
${\bf M}{\bf M}^{\dagger}$ (here ${\bf M}$ is the matrix in  (\ref{massmatrix})), 
{\it e.g.} for up-quarks, denote the normalized eigenstates as
\begin{equation}
        |up> = \sum_i \alpha_i |u_i> +  \sum_i \beta_i |u_{H_i}> 
\end{equation}
To obtain a quark eigenstate with mass $m_q$, choose 
\begin{equation}
\label{betamix}
\alpha_i \approx 1 ~~~~~~~~~~~~~~~~ \beta_i \ge \frac{v_L \Lambda_{LR}}{ v_R^2 + \Lambda_{LR}^2- m_q^2 }
\end{equation}
In the limit $\beta_i \to 0$, the eigenvalues are dependent solely on the tiny Yukawas and
 such masses correspond to the lighter quarks $m_{u,d,s,c}$. For the heavier quarks $m_{b,t}$ 
the $\beta_i$ of the above magnitudes must be carefully selected taking into consideration the
matrix structures in (\ref{massmatrix}). We obtain fits for $\beta_i$ as high as $O(10^{-2})$. This
level of mixing with exotic quarks is completely within the bounds set by unitarity of the CKM.  
Furthermore the measured entries of the CKM matrix can be matched within their error-bars by adjusting the small 
entries in Quadrant I of the mass matrices (see Appendix).

Neutrino data now seems to favor the Large Mixing Angle (LMA) solution to $\nu$-oscillation\cite{neutexp}. If 
this solution is correct, the lepton-neutrino mixing matrix ${\bf U_{MNS}}$
 (the analog\cite{MNS} of the CKM matrix for quarks)
which mixes $(e,\mu,\tau)$ with $(\nu_e,\nu_{\mu},\nu_{\tau})$ takes the form \cite{neut_matrix}
\begin{equation}
\label{nu_mix}
{\bf U_{MNS}} =  \left(
\begin{array}{ccc}
        0.7 & -0.7 & < 0.2 \\
        0.5  & 0.5  & -0.7 \\
        0.5  & 0.5  & 0.7 \\
\end{array}
 \right)_{l^{+}\nu} 
\end{equation} 
The mass-splitting between the neutrino mass-eigenstates  $(\nu_{1},\nu_{2},\nu_{3})$ is also 
constrained \cite{neut_matrix}:
\begin{equation}
\label{nu_mass}
\begin{array}{c}
\Delta m^{2}_{32} \approx 3 \cdot 10^{-3} ~~ eV^2 \\
\Delta m^{2}_{21} \approx 5 \cdot 10^{-5} ~~ eV^2 \\
\end{array}
\end{equation}
As in the quark sector, we encounter no difficulty reproducing this data with the given form of the
neutrino mass matrix, making a suitable ansatz (see Appendix) for the masses of the observable neutrinos
 in the model. Note that only a tiny amount of mixing
with exotics is necessary to drive the standard neutrino
masses to very small values (replace $\Lambda_{LR}$ with $M_{GUT}$ 
and $m_q$ with $m_\nu$ in (\ref{betamix})), leaving them the  freedom
 to mix near maximally with each other as in (\ref{nu_mix}). This 
model thus naturally predicts that mixing among the neutrinos is larger than among the quarks.

\subsection{Predictions at the Next Experiments}

Perhaps the two most testable signatures of this model are the existence of Majorana neutrinos 
and the appearance of a new vector boson, $W_R$, mediating a force between right-handed
 fermion currents which exactly mirrors the known properties of the Weinberg-Salam
 weak force. The most sensitive type of experiment at present to test both of these
predictions is neutrino-less double-$\beta$ decay, the decay $N \rightarrow N' + 2 e^-$. 
The strongest limit at present is for the half life of $^{76}Ge$ \cite{doublebeta}:
\begin{equation}
\label{Ge}
\tau_{1/2} > 1.6 \cdot 10^{25} yr
\end{equation}
The decay can proceed through either standard  $W_L$ exchange (suppressed by neutrino helicity flip)
 or in the present model through the exchange of the  $W_R$  boson or Majorana neutrino.
Given the above bound on the lifetime and including present theoretical uncertainties in 
calculating nuclear physics effects, the limits on 
the mass  $m_{\nu}$ of a Majorana neutrino or $W_R$ \cite{W_R1, W_R2} are: 
\begin{equation}
\begin{array}{l}
 m_{\nu} < (few)~eV \\
 m_{W_R}    > 1.6~TeV \\
\end{array}
\end{equation}
which tells us that $ v_R/M_{GUT} < 10^{-11}$ and $v_R \ge 1.6$ TeV, consistent with
(\ref{params}). If a decay is observed in upcoming  double-$\beta$ decay experiments
 \cite{future_beta1,future_beta2,future_beta3},
 we will hopefully have more stringent tests of the current model.

As pointed out in \cite{alignment}, SUSY models which produce flavor alignment through the RGEs 
generically predict that up-squarks are heavier than down-squarks, 
 in contrast to the usual prediction of, 
{\it e.g.}, the minimal supergravity MSSM \cite{supergrav}. This follows simply from matching 
the $\theta^2 \overline{\theta}^2$ terms in (\ref{fpsol2}), yielding a condition analogous 
to (\ref{fpsol3}):
\begin{equation}
\begin{array}{l}
{m_Q}^2+ {m_u}^2 + {m_{H_u}}^2  = \xi({g_i})  \\
{m_Q}^2+ {m_d}^2 + {m_{H_d}}^2  = \xi({g_i})  \\
\end{array}
\end{equation}
The conditions for EW symmetry breaking, 
\begin{equation}
\begin{array}{l}
\abs{\mu}^2+ {m_{H_u}}^2  = b~cot\beta + {m_Z}^2/2~cos2\beta  \\
\abs{\mu}^2+ {m_{H_d}}^2  = b~tan\beta - {m_Z}^2/2~cos2\beta  \\
\end{array}
\end{equation}
and the expression for the mass of the pseudo-scalar Higgs $A^0$, 
${m_{A^0}}^2 = 2~b/sin2\beta$, give
\begin{equation}
\label{squarks}
 {\tilde{m}_u}^2 -  {\tilde{m}_d}^2 = -({m_{A^0}}^2 + {m_Z}^2)cos2\beta 
\end{equation}
Requiring that the top Yukawa coupling $y_t$ remain perturbative down to the electroweak scale forces
$tan\beta >1$ which implies $cos2\beta<0$. Under these assumptions (\ref{squarks}) states    
$ {\tilde{m}_u}^2 >  {\tilde{m}_d}^2$ as claimed. If Supersymmetry can be found and measured with precision
at the next collider experiments\cite{colliders}, (\ref{squarks}) is a quantitative prediction.

\section{Conclusions}
\label{sec:concl}

In this paper we have constructed a specific model which illustrates a solution to the
SUSY flavor problem. The RGEs run to their fixed points, furnishing low energy Yukawa matrices that
are independent of their GUT-scale values. It produces the pattern of flavor mixings in the quark sector 
consistent with experiment. Neutrino mixings and masses are also in
agreement with experiment. Flavor violations in the squark sector are small. 
The unified scale group structure, $SO(10)_L \times SO(10)_R$, is an extension of
the original Pati-Salam unification ansatz $SU(4)_L  \times SU(4)_R$, and seems to be the
 minimal arrangement possible
to achieve all of the features derived.

The predictions this model makes at the next accelerator experiments will of course include observation of 
SUSY particles, but also a right-handed current, possibly at energies as low as $1~TeV$, 
analogous to that coupling the $W$ and $Z$ particles to standard model doublets.
 Further, if enough of the the squarks are seen and 
measured, the squark mass spectrum will have a characteristic pattern of up-squarks being heavier than 
down-squarks, the opposite of conventional predictions in the MSSM.

\section*{Acknowledgments}
 This work was supported in part by the Director, Office of Science, Office
of High Energy and Nuclear Physics, Division of High Energy Physics of the
U.S. Department of Energy under
Contracts DE-AC03-76SF00098.

\newpage

\section*{Appendix}
\label{app}

\subsection*{1. Particle Representations}

\begin{table}
\caption{Particle Representations at each Energy Scale. {\it The symmetry group (left column) 
breaks due to fields (center column) getting VEVs; see  (\ref{symm}) for breaking scheme and scales.
Prior to the breaking  (left column), fields (right column) describe the matter content.
Fields which play little phenomenological role, e.g. the $\sigma_i$'s in (\ref{su5reps}) ,
 are omitted for brevity but may be determined
from the rep's in the first row of the table and $SO(10)$ branching rules (see discussion
following (\ref{su5reps}))}}
\label{breaktable}
\begin{center}
\begin{tabular}{|l|c|l|}
\hline
Symmetry & Fields getting VEVs & Matter \\
\hline
$SO(10)_L \times SO(10)_R$ &  &   $\{ \chi_L({\bf 16};{\bf1})$ ~ 
                        $\chi_R({\bf1};\overline{{\bf 16}}) \} \times~3$ \\
       &&   $\Delta_L,\Delta'_L,\Delta''_L({\bf 10};{\bf 1})$ \\
    & $\Phi({\bf 16};\overline{{\bf 16}}) ~~ 
         \overline{\Phi}(\overline{{\bf 16}};{\bf 16})$  &  
        $\Delta_R,\Delta'_R,\Delta''_R ({\bf 1};{\bf \overline{10}})$ \\
    &&   $\Theta_L,\Theta'_L ({\bf 45};{\bf 1})$ \\
    &&   $\Theta_R,\Theta'_R ({\bf 1};{\bf 45}) ~~~~~~  \Theta ~~ ({\bf 45};{\bf 45})$\\   
\hline
  $SU(5)_L \times U(1)'_L$   & & $\psi_L ({\bf \overline{5}};3;{\bf 1};0) 
        ~~ \psi_R ({\bf 1};0;{\bf 5};-3)$ \\ 
  $\times SU(5)_R \times U(1)'_R$ &&  $\Psi_L ({\bf 10};-1;{\bf 1};0) 
       ~~ \Psi_R ({\bf 1};0;{\bf \overline{10}};1)$ \\
  &   $\Sigma_L,\Sigma'_L ({\bf 24};0;{\bf 1};0) $  
  & $ \omega ({\bf 5};-3;{\bf \overline{5}};3)
 ~~ \overline{\omega} ({\bf \overline{5}};3;{\bf 5};-3) $ \\
 & $\Sigma_R,\Sigma'_R ({\bf 1};0;{\bf 24};0)$ &
   $ \Omega ({\bf \overline{10}};1;{\bf 10};-1) 
~~ \overline{\Omega} ({\bf 10};-1;{\bf \overline{10}};1)$ \\
  &  $\Sigma_{LR} ({\bf 24};0;{\bf 24};0)$  &  $ \phi_L({\bf 5};-3;{\bf 1};0)
  ~~ \overline{\phi_L}({\bf \overline{5}};3;{\bf 1};0) $ \\
  &&   $ \phi_R({\bf 1};0;{\bf \overline{5}};3)
   ~~ \overline{\phi_R}({\bf 1};0;{\bf 5};-3) $ \\
 && $ + ~ \sigma$ ~ terms (see (\ref{su5reps})) \\
  &&  $ H_L({\bf 5};-3;{\bf 1};0)$ ~~ $\overline{H_L}({\bf \overline{5}};3;{\bf 1};0) $ \\
  &&   $ H_R({\bf 1};0;{\bf \overline{5}};3) $ ~~ $ \overline{H_R}({\bf 1};0;{\bf 5};-3) $ \\
\hline
$SU(3)_L \times SU(2)_L \times U(1)_L$   
    &&  $q({\bf3};{\bf2};1;{\bf1};{\bf1};0) 
     ~~ \overline{u}_H ({\bf\overline{3}};{\bf1};-4;{\bf1};{\bf1};0)$ \\
 $\times SU(3)_R \times SU(2)_R \times U(1)_R$ 
    && $ \overline{d}_H ({\bf\overline{3}};{\bf1};-2;{\bf1};{\bf1};0) 
   ~~ l({\bf1};{\bf2};3;{\bf1};{\bf1};0) $ \\
&& $ \overline{e}_H ({\bf1};{\bf1};6;{\bf1};{\bf1};0) ~~ u_H({\bf1};{\bf1};0;{\bf3};{\bf1};4)$  \\
&$\omega ~~~ \overline{\omega} $ & $ \overline{q}({\bf1};{\bf1};0;{\bf\overline{3}};{\bf2};-1)
 ~~  \overline{l}({\bf1};{\bf1};0;{\bf1};{\bf2};-3) $ \\
&$\Omega ~~~ \overline{\Omega} $ & $ d_H ({\bf1};{\bf1};0;{\bf3};{\bf1};2)
 ~~ e_H ({\bf1};{\bf1};0;{\bf1};{\bf1};-6) $ \\ 
&& $ \phi_L ({\bf1};{\bf2};3;{\bf1};{\bf1};0)
    ~~ \overline{\phi_L}({\bf1};{\bf2};-3;{\bf1};{\bf1};0) $ \\
&& $  \phi_R ({\bf1};{\bf1};0;{\bf1};{\bf2};-3) ~~ 
      \overline{\phi_R}({\bf1};{\bf1};0;{\bf1};{\bf2};3) $ \\
&& $ H_L ({\bf1};{\bf2};3;{\bf1};{\bf1};0) 
    ~~ \overline{H_L}({\bf1};{\bf2};-3;{\bf1};{\bf1};0) $ \\
&& $  H_R ({\bf1};{\bf1};0;{\bf1};{\bf2};-3) 
  ~~ \overline{H_R}({\bf1};{\bf1};0;{\bf1};{\bf2};3) $ \\
\hline
$SU(3)_{C} \times U(1)_{L+R}$   && $q({\bf3};1/3;{\bf2};{\bf1}) 
        ~~ \overline{u}_H ({\bf\overline{3}};-4/3;{\bf1};{\bf1}) $ \\
  $\times SU(2)_L  \times SU(2)_R$ && $ \overline{d}_H ({\bf\overline{3}};2/3;{\bf1};{\bf1})
        ~~ l({\bf1};-1;{\bf2};{\bf1}) $ \\
&& $ \overline{e}_H ({\bf1};2;{\bf1};{\bf1}) ~~ u_H({\bf3};4/3;{\bf1};{\bf1})$  \\
& $\phi_R ~~~ \overline{\phi_R}$ & $ \overline{q}({\bf3};-1/3;{\bf1};{\bf2})
   ~~  \overline{l}({\bf1};1;{\bf1};{\bf2}) $ \\
& $H_R ~~~ \overline{H_R}$ & $ d_H ({\bf3};-2/3;{\bf1};{\bf1}) ~~ e_H ({\bf1};-2;{\bf1};{\bf1}) $ \\
&& $ \phi_L ({\bf1};-1;{\bf2};{\bf1}) ~~ \overline{\phi_L}({\bf1};1;{\bf2};{\bf1}) $ \\
&& $ H_L ({\bf1};-1;{\bf2};{\bf1}) ~~ \overline{H_L}({\bf1};1;{\bf2};{\bf1}) $ \\

\hline 
$SU(3)_{C} \times U(1)_Y \times SU(2)_L$ &$\phi_L ~~~ \overline{\phi_L}$ & \\
& $H_L ~~~ \overline{H_L}$  &  MSSM\\
\hline
$SU(3)_{C} \times U(1)_{EM}$ &&  SM\\ 
\hline
\end{tabular}
\end{center}
\end{table}

\newpage

\subsection*{2. Beta Functions}

To calculate the beta function expressions (\ref{betaeqn}) and  (\ref{betaeqn2}),
\begin{equation}
\beta = \frac{{g}^3}{16 {\pi}^2} \left( \sum_{i} {C_2}(R_i)d_1(R_i)\cdots d_n(R_i)
                        - 3 C_1(G) \right)
\end{equation}
\begin{equation}
\beta_{n.S.} = \frac{{g}^3}{16 {\pi}^2} \left( {2\over3} \sum_{f} {C_2}(R_f)d_1(R_f)\cdots d_n(R_f)
+ {1\over3} \sum_{s} {C_2}(R_s)d_1(R_s)\cdots d_n(R_s)
                        - {11\over3} C_1(G) \right)
\end{equation}
one needs to know the ``indices''  ${C_2}(R_i)$ ; in Table \ref{indextable} below we 
list the indices for various rep's of $SU(5)$, $SU(3)$, and  $SU(2)$ (for a more extensive
discussion, see \cite{groups}):
  
\begin{table}
\caption{Group Theory Indices}
\label{indextable}
\begin{center}
\begin{tabular}{|l|c|c|}
\hline
$G$ & $R_i$ & ${C_2}(R_i)$ \\
\hline
 $SU(5)$ & {\bf1} & 0 \\
         & {\bf5} & $1/2$ \\ 
         & {\bf10} & $3/2$ \\ 
\hline
 $SU(3)$ & {\bf1} & 0 \\
        & {\bf3} & $1/2$ \\
\hline
 $SU(2)$ & {\bf1} & 0 \\
        & {\bf2} & $1/2$ \\
\hline 
\end{tabular}
\end{center}
\end{table}

If the group under consideration is $U(1)$, the index  ${C_2}(R_i)$ is the sum of the squares
of the (normalized)  $U(1)$ charges, $\sum_i {y_i}^2$. 

We have the further rule that  ${C_1}(SU(N)) = N$. With all of this we can compute the beta
functions (see Table \ref{betatable}).
\begin{table}
\caption{First Order Beta-Function Coefficients. {\it  The notation is
defined so ${{d{g_G}}\over{d(ln\mu)}}={{\beta_G  {{g_G}^3} }\over{16 {\pi^2}}}$.
Here $\bf{s}\sim 1~TeV$ is the scale where SUSY breaks. }}
\label{betatable}
\begin{center}
\begin{tabular}{|l|c|c|}
\hline
Energy Range & $G$ & $\beta_G$ \\
\hline
 $M_R<\mu<M_L$ & $SU(5)_R$ & $19/2$  \\
               &  $SU(3)_L$ & $29/2$  \\
               &  $SU(2)_L$ & $37/2$  \\
               &  $U(1)_L$ & $38$  \\
\hline
 $\Lambda_{LR}<\mu<M_R$
               &  $SU(3)_R$ & $29/2$  \\
               &  $SU(2)_R$ & $37/2$  \\
               &  $U(1)_R$ & $38$  \\           
               &  $SU(3)_L$ & $29/2$  \\
               &  $SU(2)_L$ & $37/2$  \\
               &  $U(1)_L$ & $38$  \\
\hline
 $v_R<\mu<\Lambda_{LR}$
          &  $SU(3)_C$ & $-3$  \\
          &  $SU(2)_R$ & $2$  \\
          &  $SU(2)_L$ & $2$  \\
          &  $U(1)_{L+R}$ & $8$  \\
\hline
 $s<\mu<v_R$
          &  $SU(3)_C$ & $-3$  \\
          &  $SU(2)_R$ & $2$  \\
          &  $U(1)_Y$ & $8$  \\
\hline
 $v_L<\mu<s$
          &  $SU(3)_C$ & $-7$  \\
          &  $SU(2)_R$ & $-8/3$  \\
          &  $U(1)_Y$ & $5$  \\
\hline
\end{tabular}
\end{center}
\end{table}

In addition to knowing the beta-functions, one must also use the matching conditions at each 
threshold where symmetries change. There are two basic rules to follow when the higher energy
symmetry $G_+$ changes to $G_-$ below the threshold:
\begin{itemize}
  \item {If $G_- \subset G_+$ , then  the matching condition is $\alpha_- = \alpha_+$}
  \item {If the generators of  $G_-$ are linear combinations of the generators 
        of several of the higher energy groups (labelled by $j$), 
        ${{\bf T_-}} = \sum_{j} c_{j}{\bf T_j}$,
 then the proper matching condition for the couplings is 
$ {\alpha_-}^{-1} =  \sum_{j,k} {c_{j}}^{2} {\alpha_{j}}^{-1}$}
\end{itemize} 
With this we have the following matching conditions:
\begin{equation}
\begin{array}{lll}
 {\alpha_C}^{-1} = {1\over2}( {\alpha_{3L}}^{-1} +  {\alpha_{3R}}^{-1}) &
        at & \mu = \Lambda_{LR} \\
 {\alpha_{1,L+R}}^{-1} = {1\over2}( {\alpha_{1L}}^{-1} +  {\alpha_{1R}}^{-1}) &
        at & \mu = \Lambda_{LR} \\ 
 {\alpha_{Y}}^{-1} =  {3\over5}{\alpha_{1R}}^{-1} +  {2\over5}{\alpha_{1,L+R}}^{-1} &
        at & \mu = v_R \\ 
\end{array}
\end{equation}

\subsection*{3. Quark and Neutrino Constraints}

Let us take the CKM matrix given as
\begin{equation}
{\bf V}_{CKM} \approx 
\left(\begin{array}{ccc}
1 & -0.2 & 0.004 \\
0.2 & 1 & -0.03 \\
0.002 & 0.04 & 1 \\ 
\end{array}\right)
\end{equation}
where we are neglecting the small $CP$-violating pieces, 
and further make the assumption that the exotic sector mixes in the same way; the total
6 by 6 ``extended-CKM'' matrix (eCKM) looks like
\begin{equation}
\label{eCKM}
{\bf V}_{eCKM} = 
\left(\begin{array}{c|c}
{\bf V}_{CKM} & < 10^{-3} \\
\hline
 < 10^{-3} & {\bf V}_{CKM} \\ 
\end{array}\right)
\end{equation}
where the off-diagonal entries are bounded from above by unitarity. We assume that the exotic
quarks mix in the same proportions as the standard quarks do only for simplicity; for
a more precise fit it is necessary to treat the exotic quark mixing as a collection of free parameters.

As in the standard model, one diagonalizes the quark mass matrix (\ref{massmatrix}) by performing
unitary transformations on the left- and right-handed quarks:
\begin{equation}
\begin{array}{ll}
u_L \longrightarrow {\bf U_u}u_L  &  d_L \longrightarrow {\bf U_d}d_L\\
u_R \longrightarrow {\bf V_u}u_R  &  d_R \longrightarrow {\bf V_d}d_R\\
\end{array}
\end{equation}
Here the $u$'s and $d$'s are 6-vectors of standard and exotic quarks. 
The phenomenological constraints are
\begin{equation}
\label{phemcon}
\begin{array}{l}
{\bf V}_{eCKM} = {\bf U_u}^{\dagger}{\bf U_d}\\
{\bf \lambda_u}{\bf \lambda_u}^{\dagger} = {\bf U_u} {\bf D_u}^2  {\bf U_u}^{\dagger}\\
{\bf \lambda_d}{\bf \lambda_d}^{\dagger} = {\bf U_d} {\bf D_d}^2  {\bf U_d}^{\dagger}\\ 
\end{array}
\end{equation}
with $D_{u,d}$ being the diagonal quark mass matrices which must accomodate the measured quark masses
in (\ref{qmasses}). There is a great deal of freedom of parameters in satisfying   (\ref{phemcon}):
the matrices ${\bf \lambda_{1,2,3}}$ as well as the small ($<<1$) Yukawas in (\ref{massmatrix}) permit
a class of solutions that give flavor-basis mass matrices like
\begin{equation}
\left(\begin{array}{ccc|ccc}
I & \cdot & \cdot & II & \cdot & \cdot \\
\cdot & o(10^{-6}-10^{-5}) & \cdot & \cdot &  o(10^{-1}-10^0) & \cdot \\
\cdot & \cdot & \cdot & \cdot & \cdot & \cdot  \\
\hline
III & \cdot & \cdot & IV & \cdot & \cdot  \\
\cdot &  o(10^{0}-10^2) & \cdot & \cdot &  o(10^{8}-10^{10}) & \cdot  \\
\cdot & \cdot & \cdot & \cdot & \cdot & \cdot  \\
\end{array}\right) \times m_W
\end{equation}
for the up-sector with the 'typical' parameters (\ref{params}) of the model ($tan\beta \approx 10$),
and similarly for the down-sector.
 The tiny couplings in Quadrant I have a significant 
effect on the mass eigenvalues; a typical pattern for  ${\bf U_u}$ might be 
\footnote{we do not furnish exact numbers here since these will depend upon the choice
 of ${\bf V_{eCKM}}$ and the ${\bf V_{u,d}}$, both
of which entail not currently observable physics and therefore may follow as suits the taste of the model-builder} 
\begin{equation}
{\bf U_u} ~ \approx ~ \left(\begin{array}{ccc|ccc}
\cdot & \cdot & \cdot & \approx 0 &  \approx 0 &  \approx 0 \\
\cdot & o(1) & \cdot &  < 10^{-3} &  < 10^{-3}  &  < 10^{-3}  \\
\cdot & \cdot & \cdot &  \approx 10^{-2} &   \approx  10^{-2}  &  \approx 10^{-2} \\
\hline
 \approx 0 &  < 10^{-3} &  \approx  10^{-2}  & \cdot & \cdot & \cdot \\
 \approx 0 &   < 10^{-3} &  \approx 10^{-2} &  \cdot & o(1) & \cdot \\
 \approx 0 &  < 10^{-3} &  \approx 10^{-2} &   \cdot & \cdot & \cdot \\
\end{array}\right)
\end{equation}
The mixing between exotic and standard quarks is
 tiny, well below unitarity and FCNC bounds, yet 
sufficient to give sizable masses to the heavier $c,t$-quarks since the exotics themselves are so massive.
The matrix calculations are therefore rather sensitive to small changes, yet can accomodate
the constraints in  (\ref{phemcon}). 

The only extra bit of analysis needed for the neutrino phenomenology is the value of the masses to insert
in the analogue of ${\bf D_{\nu}}$ (the charged lepton masses ${\bf D_{l}}$ we know fairly well,
 of course). 
If the LSND experiment's results are genuine \cite{LSND}, then it may be reasonable
 to assume an average neutrino mass around $1~eV$. The rest of the work closely parallels the above
discussion for quarks, with the result that the constraints analogous to  (\ref{phemcon}) for lepton
masses and the BNS-matrix may be satisfied within this model.

\end{document}